\documentclass [aps,prL,amssymb,amsmath,twocolumn,showpacs]{revtex4-1}

\usepackage{microtype}
\usepackage{amsmath}
\usepackage{amssymb}
\usepackage{graphicx}
\usepackage{verbatim}
\usepackage{bm}
\usepackage[dvipsnames]{xcolor}

\usepackage{graphicx,epsfig,amsfonts,amssymb}
\usepackage{hyperref}
  \hypersetup{colorlinks=true,citecolor=blue}

\usepackage{bm}
\usepackage{times}
\usepackage{lipsum}

\usepackage[all]{xy}

\newcommand{\nn}{\nonumber}

\newcommand{\bk}{\mathbf{k}}

\newcommand{\bh}{\mathbf h}

\newcommand{\bn}{\mathbf n}

\newcommand{\dg}{\dagger}

\newcommand{\ve} {\varepsilon}

\newcommand{\be}{\begin{eqnarray}}
\newcommand{\ee}{\end{eqnarray}}
\newcommand{\la}{\langle}
\newcommand{\ra}{\rangle}
\newcommand{\rar}{\rightarrow}
\newcommand{\da}{\downarrow}
\newcommand{\ua}{\uparrow}

\begin{document}

\title{Emergent  topology under slow non-adiabatic quantum dynamics}

\author {Junchen Ye} 
\author{Fuxiang Li}
\email{fuxiangli@hnu.edu.cn}
\affiliation{School of Physics and Electronics, Hunan University, Changsha 410082, China}

\date{\today}

\begin{abstract}
Characterization of equilibrium topological quantum phases  by non-equilibrium quench dynamics provides a novel and efficient scheme in detecting  topological invariants defined in equilibrium. Nevertheless, most of the previous studies have  focused on the ideal sudden quench regime. Here we provide a generic non-adiabatic protocol of slowly quenching the system Hamiltonian, and investigate the non-adiabatic dynamical characterization scheme of topological phase. The {\it slow} quench protocol is realized by introducing a Coulomb-like Landau-Zener problem,  and it can describe, in a unified way, the crossover from  sudden quench regime (deep non-adiabatic limit) to adiabatic regime. By analytically obtaining the final state vector after non-adiabatic evolution, we can calculate the time-averaged spin polarization and the corresponding topological spin texture. We find that the topological invariants of the post-quench Hamiltonian are characterized directly by the values of spin texture on the band inversion surfaces. Compared to the sudden quench regime, where one  has to take an additional step to calculate the {\it gradients} of spin polarization, this non-adiabatic characterization provides a {\it minimal} scheme in detecting the topological invariants.  
Our findings are not restricted to 1D and 2D topological phases under Coulomb-like quench protocol, but are also valid for higher dimensional system or different quench protocol. 
 \end{abstract}


\date{\today}

\maketitle


%
\section{Introduction}
The last two decades have witnessed a series of breakthroughs of discovering topological phases in various materials \cite{Hasan2010, qi2011}. Topological quantum phases are characterized by a bulk topological invariant and accompanying protected boundary modes, which are beyond the scope of the traditional Landau-Ginzburg-Wilson framework in the language of local order parameter and spontaneous symmetry breaking \cite{Landau}. Extensive studies have been performed on the characterization and detection of topological phases, not only in condensed matter physics \cite{Hasan2010, qi2011,  Chiu2016}, but also in photonic systems \cite{Haldane2008, Rechtsman2013, Khanikaev, Ozawa}, and ultracold atoms \cite{Aidelsburger2013,  Miyake2013, Liu2013, Liu2014, Jotzu, Aidelsburger2015, Wu2016}.  Great success has been achieved in experimental  detection of new topological phases by using transport measurement and angle resolved photoemission spectroscopy \cite{Konig, Hsieh, Xia09, Chang13, Xu2015, Lv15}. However, some of these experimental outcomes are not  directly related to the nonlocal property of topological phases. 

Recently, due to the improvement of experiment technique in controlling and manipulation of quantum system, great interest has been paid to the novel  dynamical  topological properties in non-equilibrium dynamics  both in theory \cite{Caio, Hu2016, Wang2017, Flaschner, Song2018, McGinley,  Lu1911, Xie2020, Hu2020, Chen2020, Qiu2019} and experiment \cite{Sun2018, Wang2019PRA, Yi2019PRL, Song2019Nat, Ji2002,  Xin2001, Niu2001}. In particular, in a series of papers of Liu's group \cite{Zhang2018Sci, Zhang2019PRA99, Zhang2019PRA100, Zhang1903, Yu2004}, a new dynamical classification theory of topological phases are proposed, in which they uncover a bulk-surface duality, stating that the $d$-dimensional ($d$D) gapped topological phase of generic multibands can reduce to a ($d-1$) D invariant defined on the so-called band inversion surfaces. Further they show that by quenching the system across the phase boundary, one can determine the $d$D topological invariant from the ($d-1$)D topological spin texture through time-averaged spin dynamics on the BIS. This idea has been generalized to characterizing topological phases in Floquet system, and systems under noisy environments.  

Most of these non-equilibrium characterization schemes, however, are  limited to the sudden quench regime, during which, the abrupt change of system Hamiltonian sets a timescale that is much shorter than all the other characteristic timescales, like the decoherence time, of the system itself.   For a more general non-adiabatic dynamics of a driven system, what would happen if the quench time is comparable to the characteristic timescale of the system? More importantly, non-adiabatic dynamics are usually thought to be detrimental to the system, since it inevitably produces excitations and topological defects, when the thermodynamic system is driven through a critical point of either classical or quantum phase transition. The density of defects follow a universal scaling law determined by the critical and dynamical exponents of the system,  which is the celebrated Kibble-Zurek mechanism \cite{Kibble, Zurek, Damski, Dziarmaga, Campo, Keesling}. In particular, methods like shortcuts to adiabaticity were proposed and realized, in order to manipulate quantum systems in timescales shorter than decoherence time, but at the same time suppressing the non-adiabatic excitations \cite{Chen10short, Odelin}.  Under these considerations, we would ask a fundamental question, is there still a non-adiabatic dynamical characterization of topological quantum phase? 

In this paper, we try to answer this question by considering a generic topological system {\it slowly} quenched from deep topologically trivial phase to nontrivial phase, and study the topological spin texture formed by the time-averaged spin polarization. Specifically, we introduce a generalized Landau-Zener (LZ) problem with a Coulomb like time-dependent term, $g/t$, in the Hamiltonian. This slow quench protocol represents a crossover from the sudden quench regime (deep non-adiabatic regime) to adiabatic regime. By analytically solving the time-dependent Hamiltonian we can determine not only the transition probability, but also the final state vector after non-adiabatic transition, from which the time-averaged spin polarization can be calculated. 
 Compared to the sudden quench regime, where one has to calculate the gradients of spin polarization along BIS, we show in this paper that non-adiabatic dynamics provides a more efficient characterization scheme of topological phase by directly measuring the values of spin polarization itself.  Our conclusions are not restricted to 1D and 2D topological phases, but can be generalized to higher dimensional system. We also show that our results are still valid for other different slow quench protocols. %
 
The paper is organized as follows. In Sec.~\ref{sec:LZ} , we introduce and solve the generalized two-level Landau-Zener problem and explain its physical implications by studying the related spin dynamics. Then in Sec. III, we present analytical results of the non-adiabatic dynamical characterization  scheme in for 1D and 2D topological phases. In Sec. IV, we further illustrate that our scheme is universal by quenching the system along different axises and with different slow quench protocol. In Sec. V, we extend our discussion to higher dimensional cases. Sec. VI is the conclusions.  

\section{generalized two-level Landau-Zener problem \label{sec:LZ}}
Since the Hamiltonian for 1D and 2D topological phases can always reduce to a direct product of independent subsystems of two-level Hamiltonian spanned in the Bloch momentum space $\bk$, we consider in this section a generalized two-level Landau-Zener problem governed by the following Coulomb like time-dependent Hamiltonian \cite{LandauQM, sinitsyn-14pra}: 
\be
H(t) = \left(\begin{array}{cc}
  g/t + \varepsilon \cos\theta  &  \varepsilon \sin\theta e^{- i\varphi}   \\
 \varepsilon \sin\theta e^{i \varphi}      &  -(g/t + \varepsilon \cos\theta)
\end{array}
\right) \label{eq:LZ}
\ee
Here $\ve$, $\theta$, and $\phi$ are parameters that describe a most generic two-level  Hamiltonian at infinite time. The essential feature of this Landau-Zener Hamiltonian is the Coulomb like time-dependent term $g/t$, with parameter $g$ determining the quench time. Parameter $g$  varies from $0$ to $\infty$, corresponding to a continuous crossover from the sudden quench limit ($g=0$) to adiabatic limit ($g\rar \infty$). The instantaneous eigen-energies can be obtained by diagonalizing the Hamiltonian (\ref{eq:LZ}), as illustrated in Fig.~1(a) as functions of time for different values of $g$. The evolution of state vector  $|\psi(t)\ra$ is governed by the time-dependent Schr${\rm\ddot{o}}$dinger equation: $i\hbar \frac{d}{dt}|\psi(t)\ra = H(t) | \psi(t)\ra$. In the following, we will set $\hbar =1$.

One reason we choose the Hamiltonian (\ref{eq:LZ}) is that this model is exactly solvable in the sense that one can express the transition probability from initial state to each final eigen-energy state in terms of simple analytical functions of parameters. More importantly, the form of $g/t$ enables us to quench the system from initially trivial Hamiltonian at $t=0^+$ to an arbitrary final state at $t=\infty$ by varying the parameters $\ve$, $\theta$ and $\varphi$, and at the same time obtain an analytical transition probability. Here in our present problem, to study the dynamical evolution of the topological spin texture, we  need to further find the transition amplitudes, which include the relative phase between the two final eigenstates. Specifically, under Hamiltonian (\ref{eq:LZ}), our task is to look for the final state vector $|\psi(\infty)\ra$ at time $t\rar \infty$ starting from the initial state $|\psi(0^+)\ra$ at $t\rar 0^+$.  At initial time $t\rar 0^+$, the time-dependent term $g/t$ is infinitely large compared with other parameters, and thus the two instantaneous eigen-energies are given by $E_{\ua, \da}=\pm g/t$ which are infinitely separated from each other, with eigenstates $\left|\ua\right\ra = (1, 0)^T$ and $\left|\da\right\ra=(0, 1)^T$, respectively. At final time $t\rar \infty$, the time-dependent term $g/t$ goes to zero, and the final Hamiltonian $H(t\rar \infty)$ has two instantaneous eigen-energies $E_{\pm}=\pm \ve$ with eigenstate vectors: $$ |+\ra =(\cos\frac{\theta}{2} e^{-i\varphi}, \sin\frac{\theta}{2})^T$$ and $$|-\ra = (\sin\frac{\theta}{2} e^{-i\varphi}, -\cos\frac{\theta}{2})^T,$$ respectively. If we prepare the system in its initial ground state, $|\psi_i \ra = (0, 1)^T$, the system will undergo a non-adiabatic transition during the evolution, and finally at time $t\rar \infty$, the system will stay not only on the final ground state $|-\ra$, but also on the excited state $|+\ra$ with LZ probability $P$.  One can see this non-adiabatic transition from Fig.~1(b), where we plot the occupation probability of state vector on the two instantaneous eigen-energy levels. One start with initial ground state on $\left|\da\right\ra$ with unit probability, and the afterward state vector would not remain on the same instantaneous ground energy level as illustrated in Fig.~1(a), but has some transition probability P of jumping onto the excited energy level.   Actually the transition probability $P$ has already been obtained by applying the no-go constraints discovered in Ref.~\cite{sinitsyn-14pra}:
\be
P=\frac{e^{-2\pi g \cos\theta} - e^{-2\pi g}}{e^{2\pi g} - e^{-2\pi g}}
\label{eq:transition probability}
\ee
which is independent of parameters $\ve$ and $\varphi$. We also present a direct calculation by solving the differential equation as in Appendix.

To understand the physical meaning of probability (\ref{eq:transition probability}), we note that $g$ denotes the time scale of driving. In the limit of large $g$, the Hamiltonian changes sufficiently slowly, and the transition probability $P \sim \exp(-2\pi g(1+\cos\theta)) \sim 0$ if $|\cos\theta| <1$, which means that the final state $|\psi(\infty)\ra$ remains on the ground state $|-\ra$,  corresponding to the adiabatic limit. In the limit  $g\rar 0$, it  corresponds to the sudden quench regime, and the initial state vector has no time to respond to the sudden change of Hamiltonian. According to the principles of quantum mechanics, the afterward evolution will be given by 
\be
|\psi(t) \ra_{sq}  =  e^{-i\ve t}c_+  |+\ra +  e^{i\ve t} c_- |-\ra,
\ee 
with the coefficients given by $c_{\pm} = \la \pm | \psi_i\ra $. Specifically, if initial state vector is in the ground state $\left| \da \right\ra=(0, 1)^T$, then $c_+ = \sin\frac{\theta}{2}$, corresponding to the transition probability in this limit $P\rar |c_+|^2$ as $g\rar 0$.  Varying $g$ from $0$ to $\infty$, the system experiences a transition from  sudden quench limit to adiabatic limit with  non-adiabatic dynamics in between, as illustrated in Fig.~1(c). 

\begin{figure}[htbp]
	\centering
	\epsfig{file=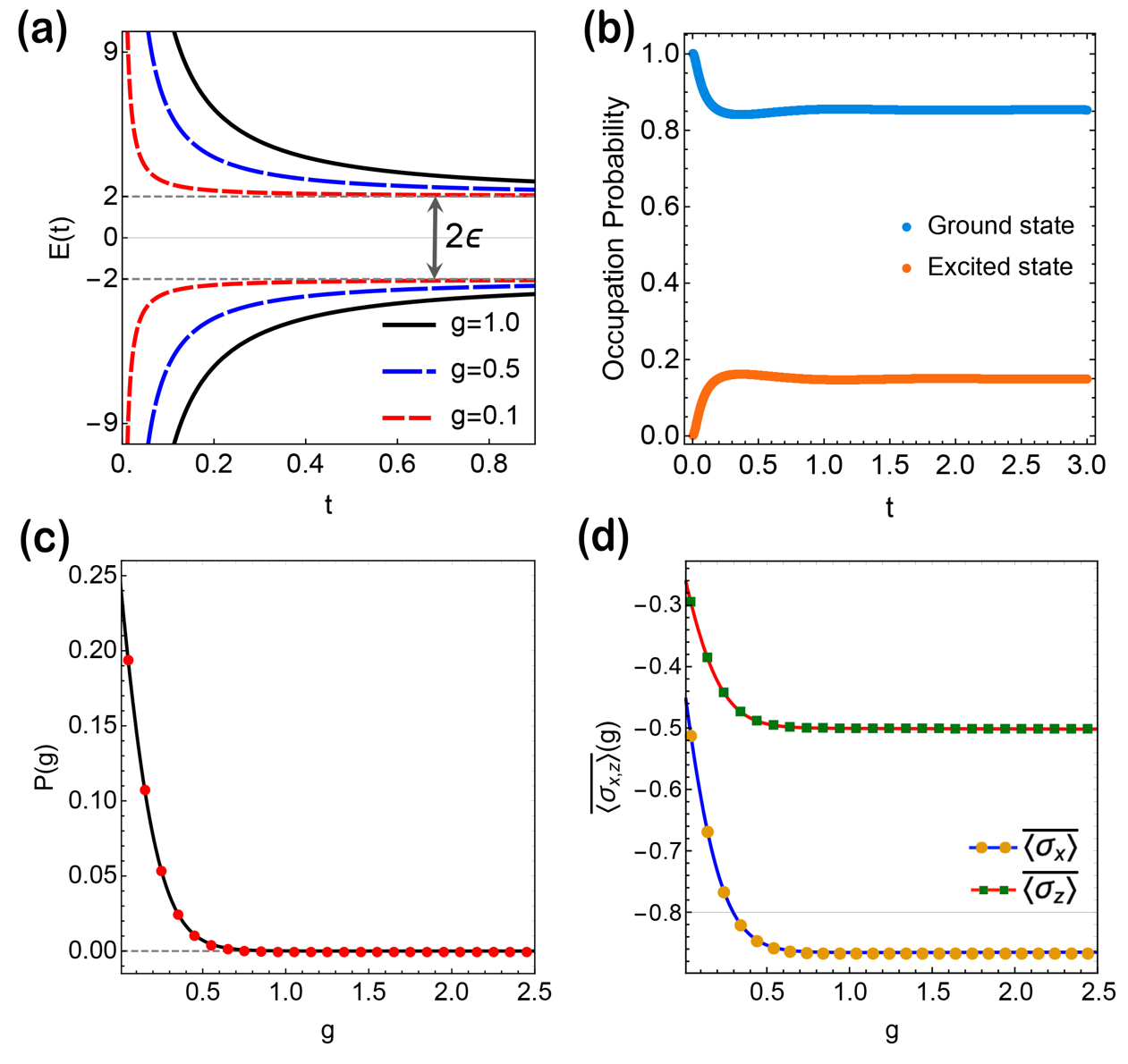, width=3.4in}
	\caption{The generalized two-level Landau-Zener problem (\ref{eq:LZ}).   (a) Instantaneous eigen-energies as functions of time $t$ (arbitrary unit) for different quench time $g$. Other parameters: $\varepsilon=2$, $\varphi=0$, and $\theta=\pi/3$, and the same for (b-d). (b) The occupation probability of time-dependent state vector $|\psi(t)\ra$  on the two instantaneous eigenstates  as the initial ground state $\left|\da \right\ra$ evolves along the red dashed energy level  in (a). (c)Transition probability $P$ from initial ground state to final excited state as a function of $g$. When $g$ increases to $1$, probability $P$ becomes zero, indicating that the system is in the adiabatic limit and the initial ground state vector remains on the ground state. The red points are numerical results and black line is the analytical result given by Eq. (\ref{eq:transition probability}). (d)  Time-averaged spin polarizations as  functions of varying $g$. Points are numerical results and the solid lines are plotted by Eq.(\ref{eq:spin polarization}). }
	\label{fig:LZ}
\end{figure}

In Appendix, we directly  solve the time-dependent Schr${\rm \ddot{o}}$dinger equation and obtain the above transition probability $P$. More importantly, we also obtain the final state vector $|\psi(t) \ra$ at long time limit, in terms of the superposition of the two instantaneous eigenstates of the final  Hamiltonian. Up to a total phase factor, we show that the final state vector oscillates with a period given by the energy gap $T=\pi/\ve$: 
\be
|\psi(t) \ra = \sqrt{P}  e^{-i\ve t} |+ \ra +  \sqrt{1-P} e^{i\ve t + \phi_0} |{}-\ra
\ee
where $\phi_0$ is a relative phase factor that is independent of time. 

Based on this solution, one can study the dynamics of spin polarization $\la {\bm \sigma}(t) \ra \equiv \la \psi(t) | {\bm \sigma}| \psi(t)\ra$ and its average $\overline{\la {\bm \sigma}\ra}$ over a period $T$.  One can treat the Hamiltonian (\ref{eq:LZ}) as $H(t) = {\bm \sigma}\cdot \bh(t)$ with $\bh(t)$ a time varying effective field 
\be
&&h_z = \frac{g}{t}+ \ve \cos\theta, \\
&&h_x=\ve \sin\theta \cos\varphi, \\
&&h_y = \ve \sin\theta \sin\varphi.
\ee
One can see that at initial time $t\rar 0^+$, the effective field is infinitely large and directs to $z$ axis, and afterwards it gradually reorients to the direction $\bn = (\sin\theta \cos\varphi, \sin\theta \sin\varphi, \cos\theta )$ with magnitude $h=\ve$. 
 To consider the spin dynamics, one can treat the spin classically, following the Bloch equation derived from the Heisenberg equation of motion:
 \be
 \frac{d}{dt} {\bm \sigma}(t) = -2{\bm \sigma }(t)\times \bh(t). 
 \label{eq:sigmat}
 \ee

In the sudden quench regime, the field $\bh(t)$, from an infinitely large vector in $z$ direction, suddenly changes to be a constant vector $\bh = \ve \bn$, while the  spin, initially polarized to $-z$ direction, would precess around $\bh$ with the Larmor frequency $2\ve$. Therefore, the time averaged spin polarization would simply be in the opposite direction of $\bh$:
 \be
 \overline{\la {\bm \sigma}\ra}_{sq}  = - \bn \cos\theta   \label{sigma_sq}
 \ee
The  magnitude $\cos\theta$ is simply the projection of the initial spin on the direction of $\bn$. 

In the adiabatic limit with large $g$, the field varies sufficiently slowly, much slower than the precession frequency induced by the field. The result is, the spin would always (adiabatically) stay in the opposite direction of changing $\bh(t)$. In this adiabatic limit, the averaged spin polarization would be 
\be
\overline{\la {\bm \sigma}\ra}_{ad}  = - \bn  \label{eq:sigma_ad}
\ee
Here the subscript $ad$ denotes `adiabatic' limit.
\begin{figure}[htbp]
	\centering
	\epsfig{file=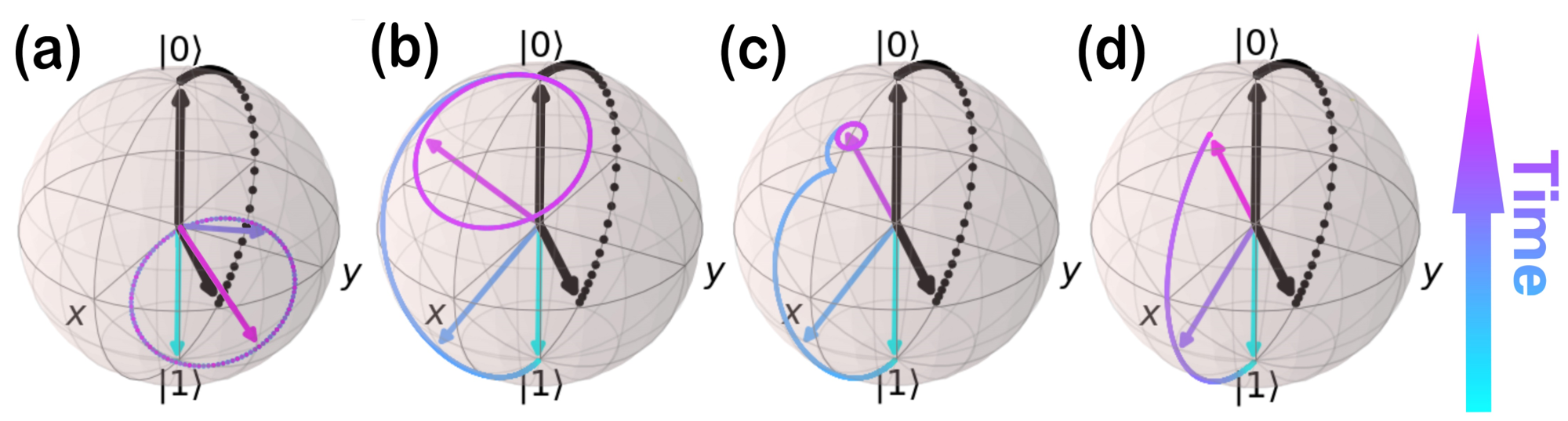, width=3.45in}
	\caption{Illustration of the crossover from sudden quench regime  to adiabatic regime  by four Bloch spheres  expressing the dynamics of spin vector given by (\ref{eq:sigmat}).  Spin polarization $\la {\bm \sigma}(t) \ra$ (colored arrow) and the unit vector of the time-dependent  effective field ${\bm n}(t)$ (black arrow) are shown with evolving time. (a) Sudden quench is approximated with $g=0.025$; (b) and (c) Non-adiabatic dynamics with different quench time, $g=5$ (b) and $g=10$ (c). (d)Adiabatic limit plotted with $g=100$. }
	\label{fig:bloch_sphere}
\end{figure}

For the general case in between the two limits, the spin would precesses around the instantaneous field direction which is also changing with a rate comparable  to the precession frequency. The final resultant averaged spin polarization would still along with or opposite to $\bn$, but with a magnitude that is between Eq.~(\ref{sigma_sq}) and (\ref{eq:sigma_ad}). A straightforward calculation shows that:
\be
 \overline{\la {\bm \sigma}\ra} = -(1-2P)\bn.
 \label{eq:spin polarization}
\ee 
This result contains not only the two limiting cases of sudden quench regime and adiabatic regime, but also describes the crossover between the two limits.

\section{quench hamiltonian}
Now we turn to the slow quench dynamics of 1D and 2D topological quantum phases, and study the emergent topological structure in the momentum space. In a series of papers \cite{Zhang2018Sci, Zhang2019PRA99, Zhang2019PRA100, Zhang1903, Yu2004}, Liu {\it et al} identified a so-called bulk-surface correspondence in the momentum space, which states that the topological invariant that characterizes a generic $d$D gapped topological phase is  equivalent to the ($d-1$)D invariant defined on the so-called band inversion surface (BIS). Based on this finding, they further proposed that the BIS and the topological (pseudo)-spin texture can be captured by the time-averaged spin polarization by suddenly quenching the system from trivial to non-trivial phase. Let's consider, for convenience, the following generic Hamiltonian spanned by $\gamma$ matrices: 
\be
{\cal H }(\bk) = {\bh}(\bk)\cdot {\bm \gamma} = \sum_{i=0}^{d} h_{i}(\bk) \gamma_i
\ee
 Here, if $d=1$ or $2$, the $\gamma$ matrices reduce to Pauli matrices ${\sigma}_j$ ($j=x, y, z$), and the Hamiltonian describes a two-band model for the topological states, including the SSH model for 1D and the Haldane model for 2D Chern insulator. Higher dimensional topological phases would require higher dimensional representation matrices of the Clifford algebra, and will be discussed in Sec.\ref{sec:high}.  
 
The BIS is defined in the Bloch momentum $\bk$ space by  arbitrarily choosing a component, say $h_0(\bk)$, of  Zeeman field vector $\bh(\bk)$, and setting it  to zero $h_0(\bk)=0$.  The remaining components of $\bh(\bk)$ are dubbed `spin-orbit' (SO) field $\bh_{so}(\bk) = (h_1, \ldots, h_d)$. The bulk-surface correspondence states that, the topological invariant $\nu(h_0, \bh_{so})$ characterizing the $d$D topological phase  is equivalent to the invariant $\omega_{d-1}(\bh_{so})$ defined on the ($d-1$)D BISs, i.e.,  $\nu(h_0, \bh_{so})=\omega_{d-1}(\bh_{so})$. Therefore, one can obtain the $d$D topological invariant by investigating the topological (pseudo)-spin texture on the ($d-1$)D BISs.

To achieve this purpose, Liu {\it et al} proposed that  the information of topological structure of spin-orbit field $\bh_{so}$ can be obtained by suddenly quenching the system from topologically trivial phase to non-trivial phase, and then the BISs can be determined by calculating the time-averaged spin polarization \cite{Zhang2018Sci, Zhang2019PRA99, Zhang2019PRA100, Zhang1903, Yu2004}. Specifically, one can choose the quenching axis to be along $h_0$, and the quenching field is set to be sufficiently large, $h_0 \gg |\bh_{so}(\bk)|$. The initial spin polarizes along the opposite direction of $h_0$. After the sudden quench, the spin would precess around the post-quench Zeeman field $\bh(\bk)$.   On the BISs with $h_0(\bk)=0$, the initial spin polarization is perpendicular to the post-quench field $\bh(\bk)$. Therefore, on BISs, the  time-averaged  spin polarization would simply vanish: $ \overline{\la \gamma_i(\bk)\ra}=0$ with $i=0, 1, 
\ldots, d$.   
This argument is in agreement with the calculation given by Eq.~(\ref{sigma_sq}), in which $\theta$ and $\bn$ are now functions of momentum $\bk$. If the quenching axis is chosen to be along $z$, then $\cos\theta(\bk) = 0$ on the BISs, leading to $\overline{\la \sigma(\bk)\ra}=0$. This means that on BISs, the time-averaged spin polarization is always zero, and one cannot obtain the topological spin texture simply from the values of spin polarization. That's why in Ref.~\cite{Zhang2018Sci, Zhang2019PRA99}, one has to  define  a new dynamical spin-texture field ${\widetilde{\bf{g}(\bk)}}$,which is the gradient of original spin texture. This additional step of calculating the topological invariants introduces  certain amount of complication in real experiments because one has to calculate the derivative of spin polarization, thus lowering the precision and efficiency.

In our paper, by introducing the non-adiabatic transitions during the slow quench dynamics, one can avoid the complication of introducing the additional gradient field. Rather, we show that one can still determine the position of BIS by $\overline{\la \sigma_z \ra}=0$ after the quench, while the remaining components of time-averaged  spin polarization are no longer zero on BISs. From them, one can directly determine the topological spin-texture that is proportional to the spin-orbit field $\bh_{so}$. 
In the following we will illustrate our findings by studying the 1D and 2D cases of topological quantum phases under non-adiabatic transition. 
\begin{figure}[htbp]
	\centering
	\epsfig{file=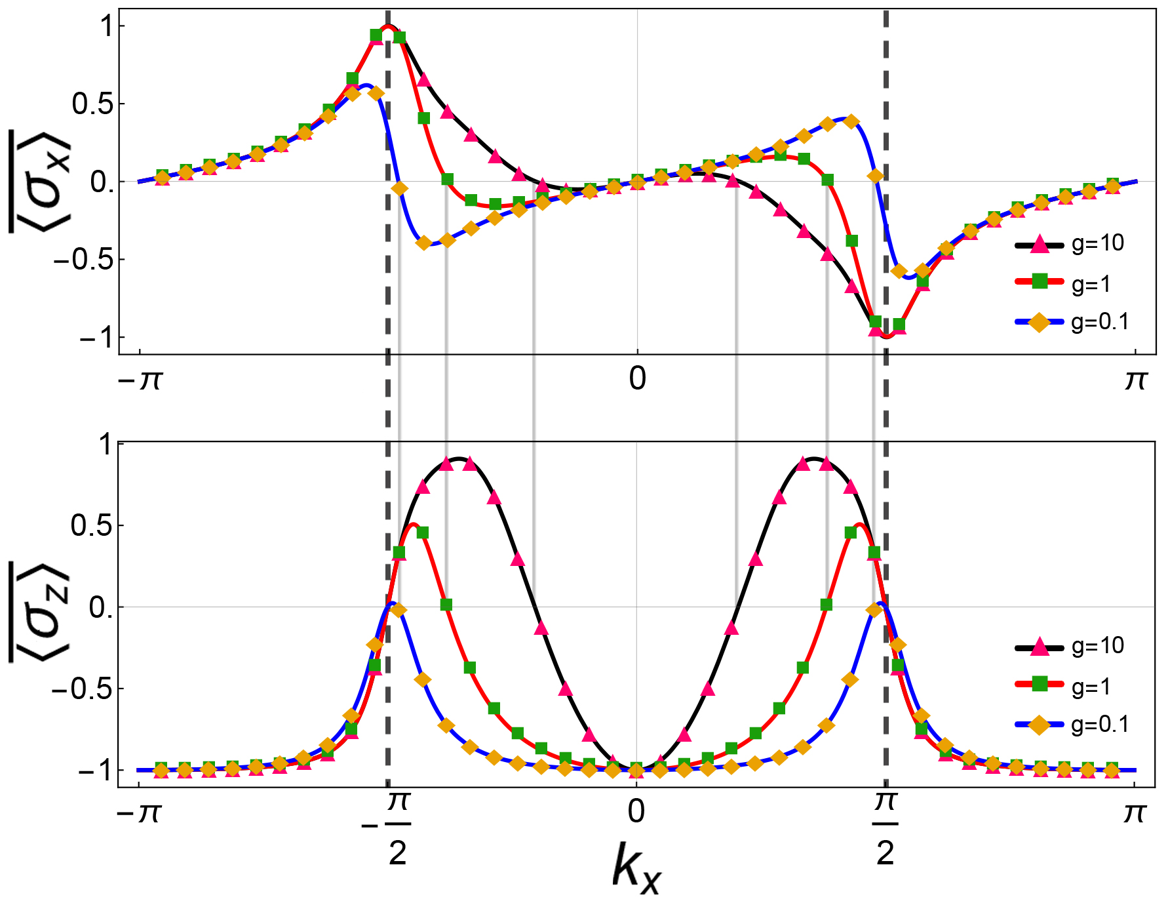, width=3.0in}
	\caption{Analytical (solid lines) and numerical (point markers) results of the time-averaged spin polarization of 1D topological model after slow quench dynamics. Three different values of driving time $g$ are plotted with $t_{so}=0.2t_{0}$ and $m_z=0$. Here we set $t_{0}=1$. The system undergoes a slow quench dynamics from time $0^+$ to a sufficiently long time $T$, and then the time average of spin components are taken over another sufficiently long time period $T$. In the numerics, $T$ is chosen to be $200$, that is already long enough compared to the largest driving time $g=10$.  Two different kinds of zeros appear in the the $z$ component $\overline{\la { \sigma}_{z}\ra}$, which are highlighted by vertical dashed lines and gray lines, respectively. While the vertical lines denote the BIS points, the gray lines denote a new kind of zeros, which we refer to as spin inversion surface(SIS), because on this surface (or points) all components of spin polarization are vanishing.  The topological spin texture can be determined by the values of $x$-component $\overline{\la { \sigma}_{x}\ra}$ on the BIS. In this case, $\overline{\la { \sigma}_{x}\ra}$ has opposite signs on the two BIS points, and give rise to a nontrivial topological structure, and thus gives the winding number $+1$.}
	\label{fig:nonlinear1D}
\end{figure}

{\it 1D case:} We first illustrate our findings by considering the 1D topological phases of AIII class with Hamiltonian spanned by two Pauli matrices ${\cal H}(k_x) = \bh(k_x) \cdot {\bm \sigma}$, and we choose the quenching axis to be along $z$ direction. 
\be
&&h_x = t_{so} \sin k_x = h_{so}, \\ 
&& h_z(t) = \frac{g}{t} + m_z- t_0\cos k_x = h_0(t)
\ee
When $h_z(t)$ is quenched from infinity at $t=0^+$, to the final value  at $t=\infty$, the system is tuned from trivial phase to a phase that depends on the value of $m_z$. If $|m_z| > t_0$, the post-quench Hamiltonian is still in the trivial phase, while if $|m_z| < t_0$, the post-quench Hamiltonian is in the topologically non-trivial phase.  The equilibrium version of this model has been realized in ultrocold atomic system \cite{Song2018}. The time changing part of $h_z$ can be precisely tuned by a bias magnetic field. 

In Fig.~\ref{fig:nonlinear1D}, we present the results of the time-averaged spin polarization of 1D topological model after slow quench dynamics. Three different values of driving time $g$ are plotted with $t_{so}=0.2t_{0}$ and $m_z=0$. The system undergoes a slow quench dynamics from time $0^+$ to a sufficiently long time $T$. After the quench, the Hamiltonian lies in the topological nontrivial regime. Then the time average of spin components are taken over another sufficiently long time $T$.  As one increases the driving parameter $g$, novel features are observed in the $z$ component $\overline{\la { \sigma}_{z}\ra}$: In contrast to the results in Ref.\cite{Zhang2018Sci} where there is only one kind of zeros, in our case, two different kinds of zeros appear in $\overline{\la { \sigma}_{z}\ra}$, which are highlighted by vertical dashed lines and gray lines, respectively. In the plots, we use the vertical lines to denote the BIS points, whose positions are given by $k_{\rm BIS}= \pm \cos^{-1}(m_z/t_0)$, and is independent of driving parameter $g$.  We use the gray lines to denote the other new kind of zeros, which we refer to as spin inversion surface (SIS), because on this surface (or points) all components of spin polarization are vanishing. The positions of SIS points, on the contrary, are dependent on parameter $g$. As one increases $g$, the separation between BIS and SIS points are enlarged.

More importantly, on the BIS points determined by $\overline{\la { \sigma}_{z}\ra}=0$, the values of $\overline{\la { \sigma}_{x}\ra}$ are no longer zero, in contrast with the case of sudden quench. The topological spin texture, therefore, can be directly determined by the values of $x$-component $\overline{\la { \sigma}_{x}\ra}$ on the BIS. In the case of Fig.~\ref{fig:nonlinear1D}, $\overline{\la { \sigma}_{x}\ra}$ has opposite signs on the two BIS points, and give rise to a nontrivial topological structure, and thus gives a winding number $+1$. Compared to the sudden quench approach in which one has to calculate the gradient of spin polarization on BIS, our non-adiabatic approach provides a more direct and simpler technique in determining the topological invariants.

These behaviors can be very well explained and understood by the formula (\ref{sigma_sq}) and (\ref{eq:spin polarization}). Indeed, if we perform a sudden quench along $z$ axis, on the BIS, the $z$-component of field is zero, i.e., $\cos\theta(\bk)=0$. This $\cos\theta(\bk)$ is nothing else but  the amplitude of the spin polarization in Eq.~(\ref{sigma_sq}), thus rendering the averaged spin polarization $\overline{\la{\bm \sigma}\ra}$ always vanishing on BIS. However, by undergoing non-adiabatic dynamics in a slow quench, the amplitude of time-averaged  spin polarization is no longer the same as $\cos\theta(\bk)$, but becomes $g$-dependent. Therefore, on the BIS, even thought $\overline{\la{\sigma}_z\ra}$  is zero, the other components would be nonzero since the amplitude in (\ref{eq:spin polarization}) is no longer zero.

\begin{figure}[htbp]
	\centering
	\epsfig{file=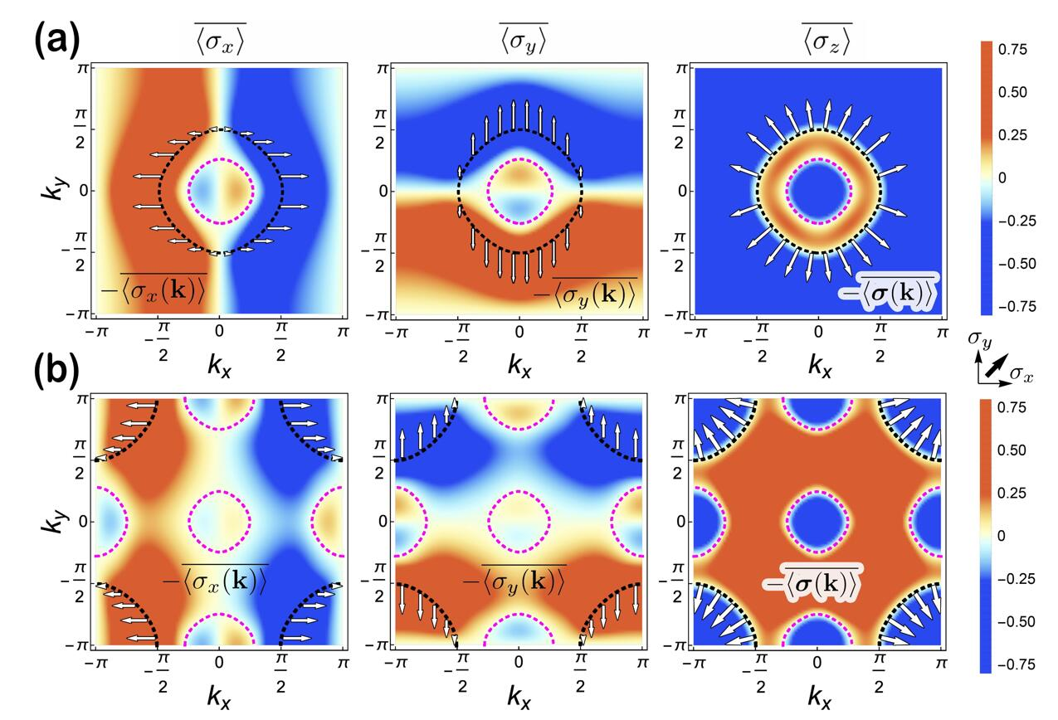, width=3.4in}
	\caption{Analytical results for 2D Chern insulators in different post-quench regime after slow quench from time $t=0^+$ to $\infty$ with effective field (\ref{eq:2Dhx}), (\ref{eq:2Dhy}) and (\ref{eq:2Dhz}), with $m_{x}=m_{y}=0$, $t_{so}^{x,y}=0.2t_{0}$. Here we set $t_0=1$ and $g=5$. (a)Time-averaged spin textures are plotted with $m_z = t_0$. SIS is the surface with all the three components vanishing $\overline{\la {\bm \sigma} \ra}=0$ (pink dashed ring) while on BIS, $\overline{\la {\sigma}_{z} \ra}=0$ but $\overline{\la {\sigma}_{x,y} \ra}\neq0$ (black dashed ring). The white arrows are the vectors formed by $-\overline{\la \sigma_{x} \ra}$ and $-\overline{\la \sigma_{y} \ra}$, indicating a nontrivial topological spin texture with  Chern number $C_{1}=-1$. (b)Time-averaged spin textures and topological spin textures $-\overline{\la {\bm \sigma} \ra}$ (white arrow) on BIS  with  $m_z = -t_{0}$. Here the BIS are located at four corners of Brillouin Zone. As the winding of $-\overline{\la {\bm \sigma} \ra}$ (white arrow) on BIS is opposite to that in (a), we identify the topological phase with the Chern number $C_{1}=1$. }
	\label{fig:nonlinear2D}
\end{figure}

{\it 2D case:} Now we generalize the above findings to the case of 2D topological states, which are generically described by a two-band Hamiltonian spanned by three Pauli matrices: ${\cal H}(\bk) = \bh(\bk, t) \cdot {\bm \sigma}$ , with the vector field given by:
\be
&&h_x=m_x+t_{so}^x \sin k_x, \label{eq:2Dhx}\\
&&h_y = m_y + t_{so}^y  \sin k_y, \label{eq:2Dhy} \\
&&h_z = \frac{g}{t}+m_z - t_0 \cos k_x - t_0 \cos k_y \label{eq:2Dhz}
\ee
The time-independent version of this Hamiltonian has been realized in recent experiment of quantum anomalous Hall effect \cite{Wu2016}. Here we want to slowly quench the $z$-component of vector field from $t=0^+$ to $\infty$ and the topological property of post-quench Hamiltonian is determined by $m_z$. For $0<m_{z}<2t_0$, the post-quench Hamiltonian describes a topological phase with Chern number $C_1=-1$, while for $-2t_0 < m_{z}<0$, the post-quench system is also topologically nontrivial but with Chern number $C_1=+1$. Analytical results of the time-averaged spin polarization can be obtained by using Eq.~(\ref{eq:spin polarization}) for each Bloch momentum $\bk$. 

 In Fig.~(\ref{fig:nonlinear2D}), we plot the three components of the time-averaged spin polarization for two different values of $m_{z}$, corresponding to two different Chern numbers of post-quench system. Similarly to the 1D case, we observe two kinds of zeros in $z$-component spin polarization $\overline{\la { \sigma}_{z} \ra}$, denoted by the black dashed ring and pink dashed ring, respectively. On the pink dashed ring, the other two components of spin polarization also vanish, and thus we name it as SIS. The black dashed ring is the BIS, given by setting the $z$-component of post-quenched vector field to be zero, i.e., $m_{z} = t_0 \cos k_x + t_0 \cos k_y$. On this BIS, the other two components of spin polarization $\overline{\la { \sigma}_{x} \ra}$  and $\overline{\la { \sigma}_{y} \ra}$ are nonzero, in contrast to the sudden quench regime. From the spin textures formed by the nonzero $\overline{\la { \sigma}_{x} \ra}$  and $\overline{\la { \sigma}_{y} \ra}$ on the BIS, as denoted by the white arrows in the contour plot of $\overline{\la { \sigma}_{z} \ra}$, one can determine the topological invariants of the post-quench Hamiltonian. For two different topological phases (a) and (b), we see that the white arrows defined on the BIS exhibit different spin textures with opposite winding numbers, thus characterizing  opposite topological Chern numbers of the post-quench 2D system.

\section{Quenching along different axis and linear quench protocol}
To complete our study of nonadiabatic transitions in the dynamical characterization of 1D and 2D topological phases , we need to further consider two cases of quench protocol, the quench along different axis and the quench with linear time-dependence. We show that our findings obtained in last section are still valid for these two cases, and the non-adidabatic dynamical characterization scheme is, therefore, universal.

\begin{figure}[htbp]
	\centering
	\epsfig{file=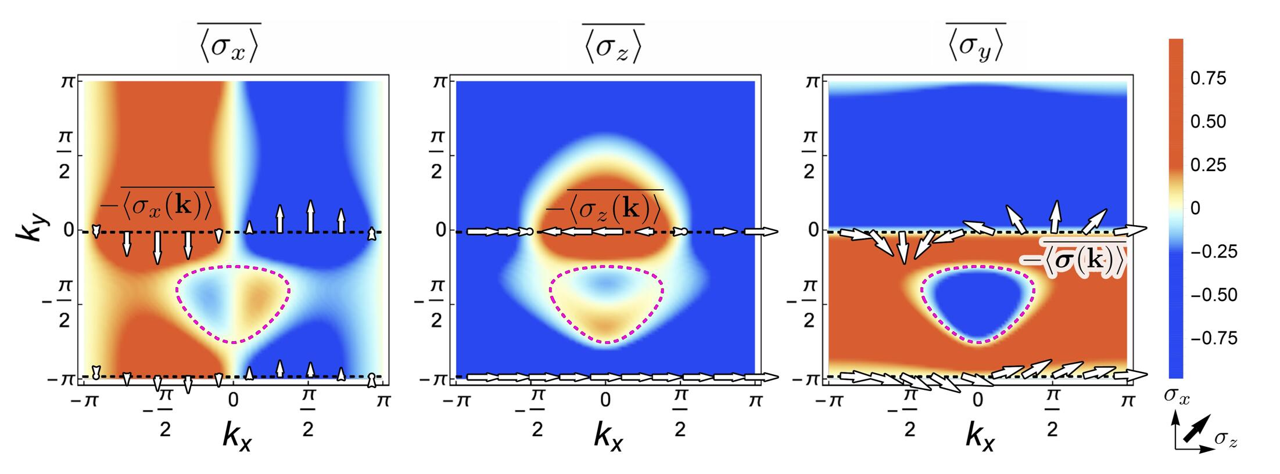, width=3.6in}
	\caption{Characterizing 2D Chern insulators by slow quench along $h_{y}({\bm k})$ axis. Time-averaged textures are analytically calculated after a slow quench under Hamiltonian $H'(t)$ in Eq.~(\ref{eq:LZy}) from  $t=0^+$ to $\infty$, with $m_{z}=t_{0}$, $m_{x}=0$, $t_{so}^{x}=0.5t_{so}^{y}=t_{0}$, and we set $t_0=1$ and $g=1$. There are two BISs (black dash line) at $k_{y}=0,-\pi$. While the winding of $-\overline{\la {\bm \sigma} \ra}$ (white arrow) is trivial along $k_{y}=-\pi$, the non-zero winding number along $k_{y}=0$ indicates the topological phase with Chern number $C_{1}=-1$. }
	\label{fig:nonlinear_qy}
\end{figure}
{\it Quench along $y$-axis.}
For the 2D case, we can also choose the quench axis to be either $x$ or $y$ directions. Let's choose $y$-axis to be the quench axis, i.e., we vary the $y$-component of vector field $h_y$ to be time-dependent: 
\be
h_y(t) = g/t +m_{y}+ t_{so}^y \sin k_y
\ee
and take other components, $h_x$ and $h_z$, to be time-independent. 
In this case, we are encountered with  a two-level Landau-Zener transition problem with the following generic time-dependent Hamiltonian:
\be
 H'(t)= \left(\begin{array}{cc}
  h_z  &  h_x-i(h_{y}+g/t)   \\
  h_x+i(h_{y}+g/t)       &  -h_z
\end{array}
\right). \label{eq:LZy}
\ee
By a time-independent unitary transformation, this Hamiltonian can be transformed to the Hamiltonian $H(t)$ (\ref{eq:LZ}). Indeed, if we choose the unitary matrix to be $U=e^{i\frac{\pi}{3}\bn_r\cdot {\bm \sigma}}$ with the unit vector $\bn_r = \frac{1}{\sqrt{3}}(1, 1, 1)$, and make a parametrization of the vector field: $h_{y}=\cos\theta$, $h_z=\sin\theta \cos\varphi$ and $h_x=\sin\theta \sin\varphi$ ,then it can be readily checked that $U^{\dg} H'(t) U = H(t)$. With this unitary transformation in hand, one can argue that the Landau Zener transition probability, defined by Hamiltonian $H'(t)$ in (\ref{eq:LZy}) from initial ground state to final excited state, is the same as that obtained in Eq.~(\ref{eq:transition probability}). The argument is presented in Appendix. The final wave function, and the time-averaged spin polarization can also be readily obtained. Using these results, we can plot in Fig.~(\ref{fig:nonlinear_qy}) the time-averaged spin polarization of the 2D Chern insulator after quench along $y$-axis. In this case,  $m_{y}$ is set to be zero, and thus the BISs are given by $\sin k_y =0$ giving rise to $k_y = 0$ or $\pi$. Again, on this BISs, the values of $\overline{\la {\sigma}_y \ra}$ vanishes, but $\overline{\la {\sigma}_x \ra}$ and $\overline{\la {\sigma}_z \ra}$ are nonzero,  from which, the topological spin texture can be straightforwardly determined. We see from the figure that, while the winding of $-\overline{\la {\bm \sigma} \ra}$ (white arrow) is trivial along $k_{y}=-\pi$, the non-zero winding number along $k_{y}=0$ indicates the topological phase with Chern number $C_{1}=-1$.

 \begin{figure}[htbp]
	\centering
	\epsfig{file=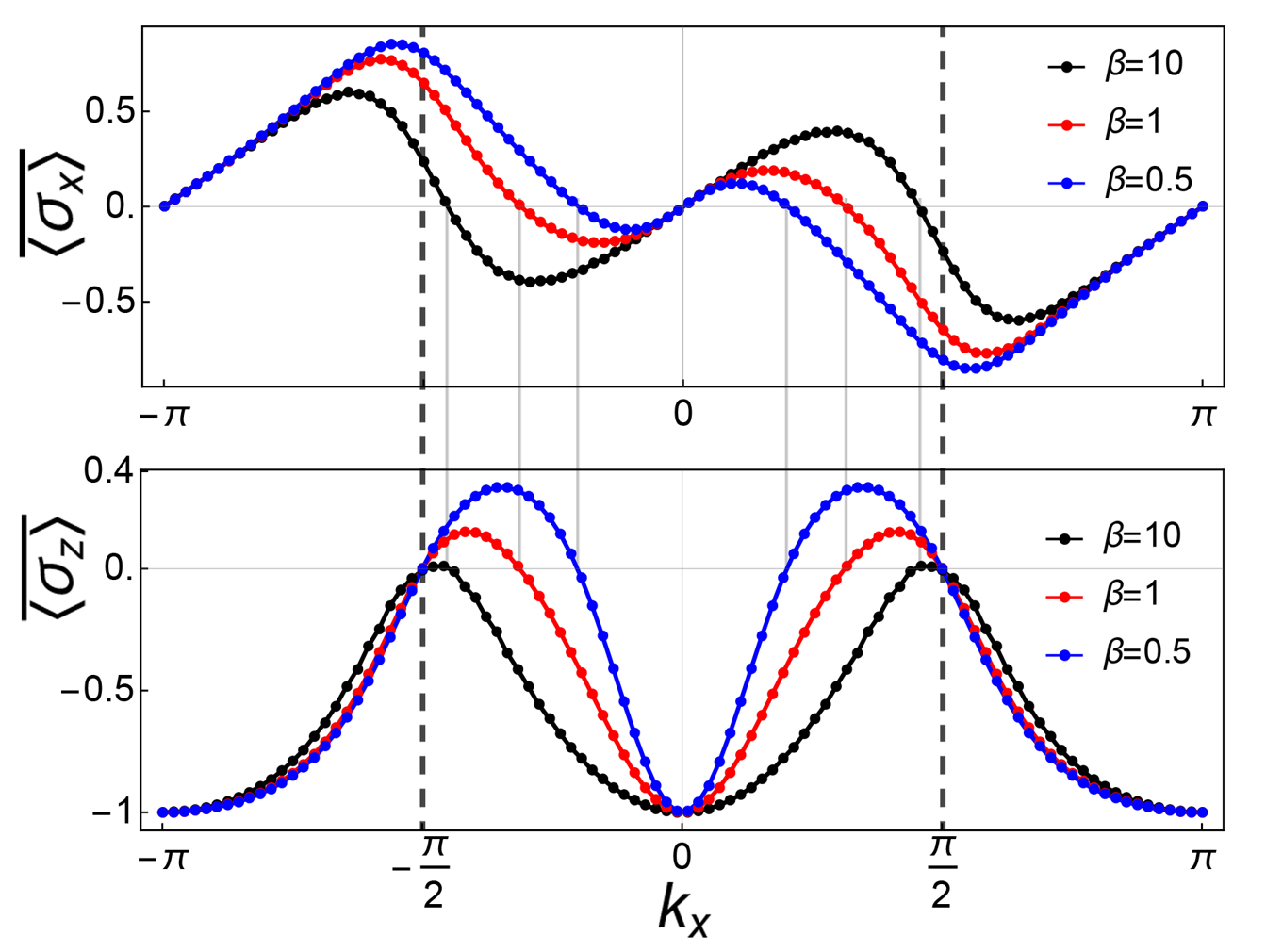, width=3.05in}
	\caption{Numerical results for the 1D topological model with linear quench protocol $\beta t$. Time-averaged components $\overline{\la { \sigma}_{x}\ra}$ (upper panel) and $\overline{\la {\sigma}_{z}\ra}$ (lower panel) of spin texture $\overline{\la {\bm \sigma}\ra}$ are shown for different values of quench speed $\beta$. The quench is taken from $t=-20$ to $t=0$ with $m_z=0$ and $t_{so}=0.6t_{0}$ so that the post-quench Hamiltonian is in topologically nontrivial phase. After the quench, the system is under free evolution and the spin polarization is averaged over a long time period. The SISs and  BISs are denoted by gray lines and  dashed lines, respectively. The opposite sign of $-\overline{\la {\bm \sigma}_{x}\ra}$ on BIS characterizes the nontrivial topology, and gives the winding number $+1$.}
	\label{fig:linear1D}
\end{figure}

{\it Linear quench protocol.} One may wonder whether our findings of non-adiabatic dynamical detection of topological phases are constrained to only the specific $g/t$ quench protocol given by (\ref{eq:LZ}). Here we want to apply a linear quench protocol to this problem, and show that our non-adiabatic approach is valid in a general sense. Actually, the only reason we adopt the $g/t$ quench protocol is that, we can obtain simple analytical expressions of transition probability and time-averaged spin polarization for an arbitrary post-quench Hamiltonian. For the linear quench protocol, one can still obtain the transition probability for arbitrary post-quench Hamiltonian in terms of special functions, but the underlying physics becomes obscured. Therefore, we will present only the results of numerical calculations. 

Specifically, for 1D case, we quench the Hamiltonian ${\cal H}(k_x) = \bh(k_x) \cdot {\bm \sigma}$ with a linearly time-dependent vector field: 
\be
&&h_x = t_{so} \sin k_x = h_{so}, \\ 
&& h_z(t) = \beta t + m_z- t_0\cos k_x = h_0(t)
\ee
Here, $\beta$ describes how fast the quench is. The quench starts from $t= -\infty$ to a finite value of $t$. As presented in Fig.~\ref{fig:linear1D}, similar features emerge as in the $g/t$ quench protocol. The zeros of $\overline{\la {\sigma}_{z}\ra}$ are characterized as BISs (black dashed lines) and SISs (gray lines), respectively. The positions of BIS points determined by $m_z = t_0 \cos k_x$  are the same as in $g/t$ protocol, while the positions of  SIS points are $\beta$ dependent. On the SISs, all the components of spin polarization vanish. On the BISs, the fixed component $\overline{\la {\sigma}_{x}\ra}$ is nonzero. Its opposite signs on the two BIS points characterizes the nontrivial topology, and gives the winding number $+1$.

\begin{figure}[htbp]
	\centering
	\epsfig{file=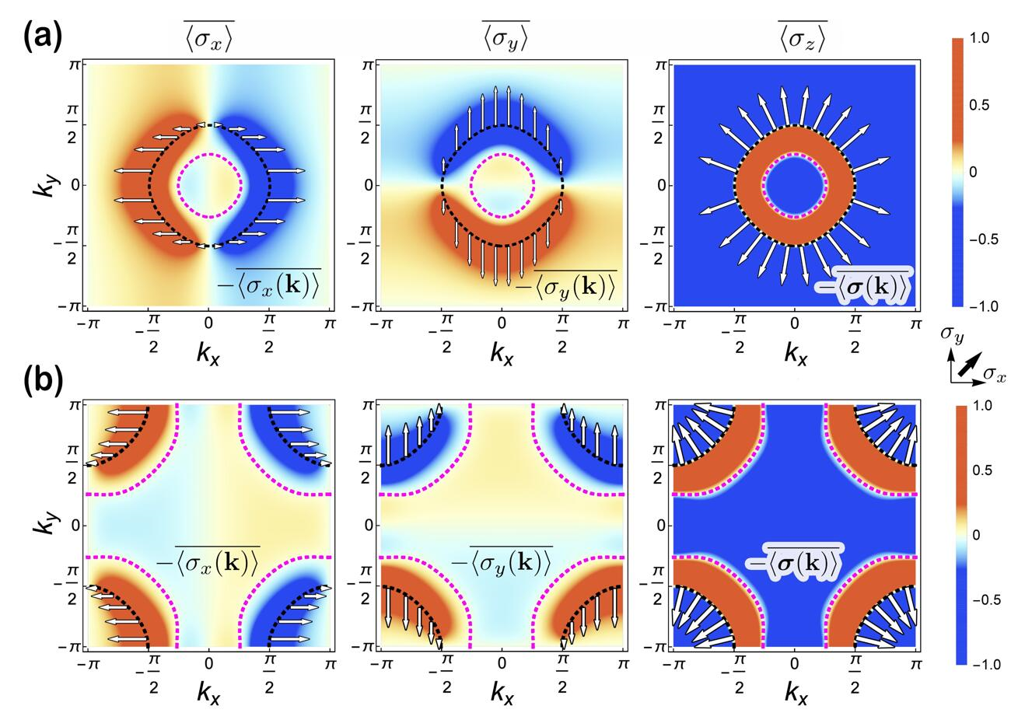, width=3.5in}
	\caption{Numerical results for 2D topological phase with different quench protocal $\beta t$, with $m_{x}=m_{y}=0$, $t_{so}^{x,y}=0.6t_{0}$, and $\beta=0.8$  by setting $t_0=1$. (a)Time-averaged spin textures are plotted after slow quench from $t=-10$ (trivial) to $t=0$ (topological) with $m_{z}= t_0$.  SIS is shown pink dashed ring with $\overline{\la {\bm \sigma} \ra}=0$  while BIS is shown by black dashed ring with $\overline{\la {\bm \sigma}_{z} \ra}=0$ but $\overline{\la {\sigma}_{x,y} \ra}\neq0$ . The white arrows denote the topological spin texture, composed of $-\overline{\la \sigma_{x,y} \ra}$, indicating  Chern number $C_{1}=-1$. (b)Time-averaged spin textures and topological characteristics $-\overline{\la {\bm \sigma} \ra}$ on BIS (white arrow) are plotted after slow quench from $t=-12$ (trivial) to $t=0$  (topological) with $m_z=-t_0$. Here the BIS are located at four corners of Brillouin Zone. From the white arrows, we identify the topological phase with the Chern number $C_{1}=1$ opposite to that in (a).}
	\label{fig:linear2D}
\end{figure}

 Now we apply the linear quench protocol to the case of 2D topological states. The vector field is quenched in the following way:
\be
&&h_x=m_x+t_{so}^x \sin k_x,~~h_y = m_y + t_{so}^y  \sin k_y, \\
&&h_z = \beta t + m_z  - t_0 \cos k_x - t_0 \cos k_y. 
\ee
 In Fig.~(\ref{fig:linear2D}), we plot the three components of the time-averaged spin polarization for two different values of $m_{z}$, corresponding to two different Chern numbers of post-quench system. The figures appear to be different from those in Fig.~(\ref{fig:nonlinear2D}), but the essential features are the same. We observe two kinds of zeros in $z$-component spin polarization $\overline{\la { \sigma}_{z} \ra}$, denoted by the black dashed ring and pink dashed ring, respectively. On the pink dashed ring (SIS), the other two components of spin polarization also vanish. The black dashed ring is the BIS.  On this BIS, the other two components of spin polarization $\overline{\la { \sigma}_{x} \ra}$  and $\overline{\la { \sigma}_{y} \ra}$ are nonzero. From the spin textures formed by the nonzero $\overline{\la { \sigma}_{x} \ra}$  and $\overline{\la { \sigma}_{y} \ra}$ on the BIS, as denoted by the white arrows in the contour plot of $\overline{\la { \sigma}_{z} \ra}$, one can determine the topological invariants of the post-quench Hamiltonian. For two different topological phases (a) and (b), we see that the white arrows defined on the BIS exhibit different spin textures with opposite winding numbers, thus characterizing  opposite topological Chern numbers of the post-quench 2D system. 
 
\section{Slow Quench of Higher dimensional phases \label{sec:high}}

\begin{figure}[htbp]
	\centering
	\epsfig{file=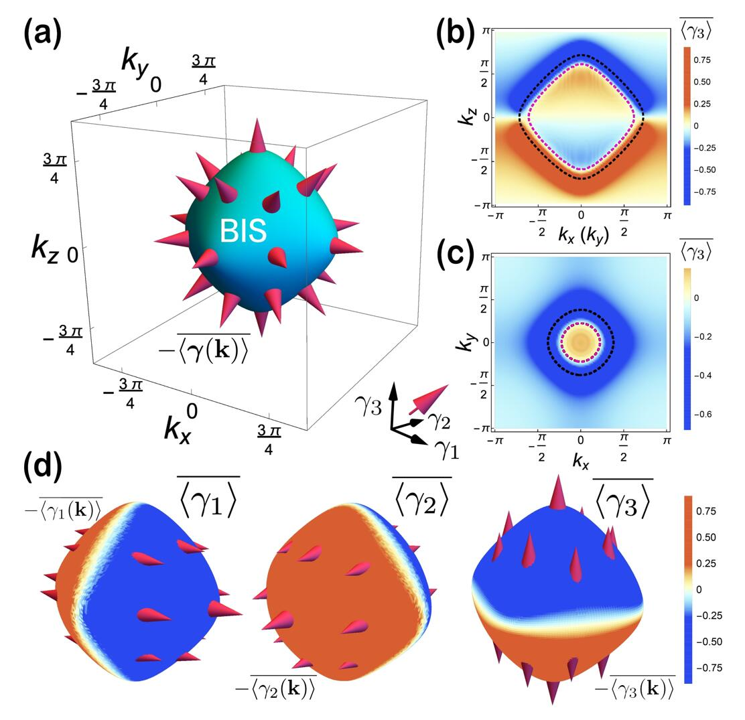, width=3.4in}
	\caption{Non-adiabatic characterization of 3D chiral topological phases. The quench is simulated numerically from $t=0.015$ to $1000$, with parameters $t_{so}=0.2t_{0}$ and $g=1$ by setting $t_{0}=1$.  (a)The BIS (sphere) defined by $\overline{\la \gamma_{0}({\bm k}) \ra}=0$ and the topological spin texture field  $-\overline{\la {\bm \gamma ({\bm k})} \ra}$ (pink arrows), composed of the three components $-\overline{\la \gamma_{1,2,3}({\bm k}) \ra}$ [pink arrows on the three spheres in (d)]. (b) and (c)are the time-averaged spin texture $\overline{\la \gamma_{3} \ra}$ on cross sections $k_{x,y}=0$ (b), and $k_{z}=\pi/2$ (c). Again, BISs and SISs are denoted by black dashed ring and pink dashed ring, respectively. (d) The three components of time-averaged spin texture $\overline{\la \gamma_{1,2,3} \ra}$ on BIS. Their values of topological spin texture components $-\overline{\la \gamma_{1,2,3}({\bm k}) \ra}$ are illustrated by pink arrows.}
	\label{fig:3D}
\end{figure}

In this section, we further study the non-adiabatic quench processes in higher dimensional topological phases, and show that our findings are not limited to 1D and 2D cases.  3D and 4D topological phases need at least a $4$-dimensional representation of Clifford algebra, and the Hamiltonian is described by four-band model. This requires us to consider a four-state Landau Zener problem, which lies in the context of multi-state Landau Zener problem.  Exactly solvable models of multi-state Landau Zener problems have been recently found to exist in many different forms \cite{sinitsyn2017integrable, sinitsyn-14pra, sinitsyn16pra, li17pra,  li18prl}. In particular, integrability conditions were found and new solvable models are discovered.  However, in our case of four-state problem, the exact solutions are still absent or it may be not exactly solvable. Therefore, we will solve the problem by numerical calculations.

The Hamiltonian describing the 3D topological phases can be written as: $H(\bk) = \sum_{j=0}^3 h_j(\bk) \gamma_j$, with
\be
&&h_0=\frac{g}{t}+ m_z -t_0 \sum_i \cos k_i \\ 
&& h_i= t_{so} \sin k_i, ~{\rm with}~ i=x, y,z
\ee
For the $\gamma$ matrices, we adopt the convention as used in Ref.~\cite{Zhang2018Sci}: $\gamma_0=\sigma_z \otimes \tau_x$, $\gamma_1 = \sigma_x \otimes 1$, $\gamma_2 = \sigma_y \otimes 1$ and $\gamma_3 = \sigma_z \otimes \tau_z$, where $\sigma_i$ and $\tau_i$ are Pauli matrices. The slow quench starts from time $t=0^+$ with large $h_0$ so that the system is initially in the topologically trivial phase. At time $t\rar \infty$, the system may lie in different phases that are dependent on the values of $m_z$: $|m_z|>3t_0$, trivial phase; $t_0<m_z <3t_0$, topological phase with winding number $\nu_3=-1$;  $-t_0 < m_z<t_0$, with $\nu_3=2$; and $-3t_0< m_z< -t_0$, with $\nu_3=-1$. In the Fig.~\ref{fig:3D}, we numerically calculate the time-averaged spin polarization after the slow quench. One can see that, on the 2D BIS, the values of three spin polarization components constitute the pink arrows that gives a nonzero Chern number $C_1=-1$, thus corresponding to 1 nontrivial topological 3D phase with winding number $\nu=-1$. 

\section{Discussions and Conclusions}
Before concluding, we argue that the non-adiabatic protocol is readily accessible in ultracold atomic experiments. The time changing part of the quenching axis can be precisely tuned by a bias magnetic field, as can be realized in experiment \cite{Song2018, Sun2018}. Furthermore, in order to observe the effect of non-adiabatic transition in the spin polarization, one needs to ensure that the non-adiabatic transition time $t_{c}$ must be much smaller than the decoherence time $\tau$ of the ultracold atomic system, which is longer than $1$ ms. The non-adiabatic transition time $t_{c}$ can be estimated by the celebrated LZ transition probability $P=\exp(-2\pi\Delta^2/(\hbar \beta))$, where $\Delta$ is the characteristic energy scale and $\beta$ is the velocity of driving field. The adiabatic limit is achieved as $\Delta^2/(\hbar \beta) \sim 1$, while at non-adiabatic transition time we have $\beta t_c = \Delta$. The two relations give rise to $t_c\sim \hbar/\Delta$, which is of the order of $10$ $\mu$s in the real experiment.  Therefore, the condition $t_c\ll \tau$ is satisfied, and the non-adiabatic effect could be observed.

It is promising and of great interests to generalize the non-adiabatic dynamical characterization scheme developed in this work to other categories of topological phases and to situations with dynamical noises. In particular, we note that, in Ref.~\cite{Zhang2020noise}, the dynamical topology under sudden quench regime was studied  in the presence of dynamical noise. They found that the dynamical topology is robust against the weak noise, and a novel dynamical topological transition is observed at a critical noise strength. We believe that our scheme is also robust against weak noise and more novel features are to be explored in the future work.


In summary, we have studied the non-adiabatic dynamical characterization of topological phase under slow 	quench protocol. A generalized Landau-Zener Hamiltonian is introduced and analytically solved, which enables us to consider the quench dynamics from sudden quench regime to adiabatic regime. 
This finding is contrary to the traditional opinion that non-adiabatic transition would usually produce  defects in the system and thus introduce complications. We find that the topological invariants of the post-quench Hamiltonian are characterized directly by the values of spin texture on the band inversion surfaces, which provides a more efficient characterization scheme than the sudden quench regime, where one  has to take an additional step to calculate the  gradients of spin polarization. 
We extend our discussions from 1D and 2D topological phases under Coulomb-like quench protocol to more general situations, like higher dimensional system and different quench protocol, which shows that the non-adiabatic classification scheme is universal. Our work also presents a different concept against the traditional opinion that non-adiabatic transition would usually produce  defects in the system and thus introduce complications.

\section*{Acknowledgements}
The authors are grateful for the helpful discussions with Xiong-Jun Liu and for his critical reading of the manuscript. This work was supported by NSFC (No. 11905054) and by the Fundamental Research Funds for the Central Universities from China. 

\section*{Appendix}
{\it The wave function after the LZ transition.} 
In this appendix, we present the details of solving the generalized LZ problem by directly integrating  the Schr${\rm \ddot{o}}$dinger equation. We first consider the following two-level Hamiltonian parameterized by $\ve$ and $\theta$ and is driven with a time-dependent term $g/t$: 
\be
H(t) = \left(\begin{array}{cc}
  g/t + \varepsilon \cos\theta  &  \varepsilon \sin\theta   \\
 \varepsilon \sin\theta     &  -(g/t + \varepsilon \cos\theta)
\end{array}
\right) \label{eq:LZ2}
\ee
The wave function is a spinor $|\Psi(t)\ra = (u(t), v(t))^T$., and they satisfy the following Schrodinger equation:
\be
&&i\frac{du}{dt} = (g/t + \ve \cos\theta) u + \ve \sin\theta v \\
&& i\frac{dv}{dt} = -(g/t + \ve \cos\theta) v + \ve \sin\theta u
\ee
Here we have set $\hbar$ to be $1$. 
From the first equation, we express $v$ in term of $u$,
and then substitute this expression into the second equation, 
\be
\frac{d^2 u}{dt^2 } + \Big[ \frac{g^2 -ig}{t^2} + \ve^2 + \frac{2g\ve \cos\theta}{t} \Big] u= 0
\ee

Making a transformation of $u(t) = t^{-ig} e^{i\ve t} \phi(t)$, and a variable change
 $t=\tau/(-2i\ve)$, we arrive at the following confluent hypergeometric equation for $\phi(t)$:
\be
\tau \phi'' + ( b-\tau) \phi' - a \phi =0
\ee
with
\be
&&a =  -ig (1+ \cos\theta), \\
&& b = -2i g
\ee
This equation is also called Kummer's equation, and it has two standard solutions $M(a, b, \tau)$ and $U(a, b, \tau)$. Their limiting forms as $\tau\rar 0$ are:  $M \rar 1$ and $U  \rar \frac{\Gamma(1-b)}{\Gamma(a+1-b)}$ with $b$ being pure imaginary for our case. 
Therefore, for the initial condition $ |u(t\rar 0)| = 1 $ and $v(t\rar 0) =0$, we must choose generally $\phi(\tau) = c_1 M(\tau) + c_2 U(\tau)$, and thus
$ u(t) = t^{-ig} e^{i\ve t} \phi(\tau)$.
The initial condition of $v=0$ leads to $d\phi/d\tau  -\cos^2\frac{\theta}{2} \phi = 0$. One is ready to check that this condition gives rise to $c_2=0$, and $|c_1| =1$. Now we can choose $c_1=1$ up to an unimportant phase factor, and the solutions become:
\be
&&u(t) = t^{-ig} e^{i\ve t} M(a, b, \tau)\nn\\
&&v(t) =  t^{-ig}e^{i\ve t} \cot\frac{\theta}{2} [M(a+1, b+1, \tau) - M(a, b, \tau)] \nn \\
\ee

 For the transition from state $(1, 0)^T$ at $\tau\rar 0$ to final state $|+ \ra = (\cos\frac{\theta}{2}, \sin \frac{\theta}{2} )^T$, we need to consider the asymptotic behavior of amplitude function 
\be
p_+ &&= \la +| \Psi(t) \ra = \cos\frac{\theta}{2} u(t) + \sin\frac{\theta}{2} v(t) \nn \\
&&= t^{-ig} e^{i\ve t} \cos\frac{\theta}{2} M(a+1, b+1, \tau)
\ee
 In the long time limit, using the asymptotic behavior of Kummer function, 
 we have
\be
p_+ = t^{-ig} e^{-i\ve t} \cos\frac{\theta}{2}  \frac{\Gamma(1+b)}{\Gamma(1+a)}  (-2i\ve t)^{a-b}
\ee
Applying the property of Gamma function, 
we can obtain the transition probability from initial excited state  to the final excited state:
\be
|p_+|^2 = \frac{e^{2\pi g} - e^{-2\pi g\cos\theta}}{e^{2\pi g} - e^{-2\pi g}}
\ee
It is the also the probability from the ground state $(0, 1)^T$ to the final ground state $1-P$ that we have used in the main text. 

Transition from initial excited state to ground state with energy $-\ve$  can also be obtained using similar procedure, and one finds
\be
p_- = t^{-ig} e^{i\ve t}\frac{1}{\sin\frac{\theta}{2}} \frac{\Gamma(b)}{\Gamma(b-a)}  (2i\ve t)^{-a}
\ee
The corresponding probability is given as:
\be
|p_-|^2 =  \frac{ e^{-2\pi g\cos\theta}-e^{-2\pi g} }{e^{2\pi g} - e^{-2\pi g}}
\ee
It is seen that total transition probability is conserved: $|p_+|^2 + |p_-|^2 = 1$. 

The most interesting part is that, for $p_+$ and $p_-$, the relative phase is oscillating with time,
\be
\delta \phi = -2\ve t - 2g \cos\theta \ln t
\ee
At large $t$, the second term is vanishingly small compared with the first term. Thus, we see that the relative phase oscillates with frequency $2\ve$. Therefore, one can write down the final wave function, up to a phase factor, in the form as: 
\be
|\Psi(t)\ra = |p_+| e^{i\delta\phi} |+\ra + |p_-| |-\ra . \label{eq:psit1} 
\ee

Now we consider the evolution under the following Hamiltonian: 
\be
H'(t) = \left(\begin{array}{cc}
  g/t + \varepsilon \cos\theta  &  \varepsilon \sin\theta e^{- i\varphi}   \\
 \varepsilon \sin\theta e^{i \varphi}      &  -(g/t + \varepsilon \cos\theta)
\end{array}
\right)
\ee
Actually, this Hamiltonian can be obtained from (\ref{eq:LZ2}) by a unitary transformation $\hat{U} =  e^{ i\varphi \sigma_z/2}$: 
\be
\hat{U}^{\dg}H'(t)  \hat{U} &&=H(t) 
\ee
Thus we reduce the problem to the previous one by formally writing down the evolution of wave function:
\be
|\Psi'(t)\ra && = \hat{\cal T}e^{-i\int^t_0 d\tau H'(\tau)} |\Psi'(0)\ra \\
&& = \hat{U} \hat{\cal T}e^{-i\int^t_0 d\tau H(\tau)}\hat{U}^{\dg} |\Psi'(0)\ra 
\ee
For the initial state being either $|\Psi'(0)\ra = (1, 0)^T$ or $(0, 1)^T$, the application of $\hat{U}^{\dg}  \equiv e^{i\phi\sigma_z/2}$ on it is equivalent to adding a universal phase factor, i.e., $\hat{U}^{\dg}|\Psi'(0)\ra = e^{\pm i\phi/2} |\Psi(0)\ra$ with $\pm$ corresponding to $|\Psi(0) \ra=  (1, 0)^T$ or $(0, 1)^T$. Thus, we build the relation between the two final states  $|\Psi'(t)\ra$ and $|\Psi(t)\ra$: $|\Psi'(t)\ra  = e^{\pm i\phi/2} \hat{U} |\Psi(t) \ra$.
From this relation, we can  express the $|\Psi'(t)\ra$ in terms of the eigenvectors of final $H'(t)$, by considering Eq.~(\ref{eq:psit1}) and the relation $|\ua'\ra = \hat{U} |\ua \ra$, $|\da'\ra = \hat{U} |\da \ra$:
\be
|\Psi'(t) \ra = e^{\pm \phi/2} (p_+ |\ua'\ra + p_- |\da'\ra ).
\ee
Thus we arrive at the conclusion that, for the new Hamiltonian $H'(t)$, the transition amplitude from the initial state $(1, 0)^T$ to final eigenstate of $H'(t \rar \infty)$ is also given by $p_{\pm}$ up to a overall phase factor.

\end{document}